\documentclass[preprint,12pt,authoryear]{elsart5p}

\usepackage{amsmath}
\usepackage{amssymb}
\usepackage{subfig}
\usepackage{graphics}
\usepackage{epsfig}

\journal{Physics Letters B}

\begin{document}
\renewcommand{\textfraction}{0.00000000001}
\renewcommand{\floatpagefraction}{1.0}
\topmargin -2.cm
\begin{frontmatter}
\title{First measurement of the polarization observable $E$ and helicity-dependent 
cross sections in single $\pi^{0}$ photoproduction from quasi-free nucleons}


\author[basel]{M.~Dieterle},
\author[basel]{L.~Witthauer},
\author[mainz]{ F.~Cividini},
\author[basel]{S.~Abt},
\author[mainz]{P.~Achenbach},
\author[mainz]{P.~Adlarson},
\author[hiskp]{F.~Afzal},
\author[regina]{Z.~Ahmed},
\author[kent]{C.S.~Akondi},
\author[glasgow]{J.R.M.~Annand},
\author[mainz]{H.J.~Arends},
\author[edinburgh]{M.~Bashkanov},
\author[hiskp]{R.~Beck},
\author[mainz]{M.~Biroth},
\author[dubna]{N.S.~Borisov},
\author[pavia]{A.~Braghieri},
\author[gdwu]{W.J.~Briscoe},
\author[pavia]{S.~Costanza\thanksref{1}},
\author[halifax]{C.~Collicott},
\author[mainz]{A.~Denig},
\author[mainz,gdwu]{E.J.~Downie},
\author[mainz]{P.~Drexler},
\author[mainz]{M.I.~Ferretti-Bondy},
\author[glasgow]{S.~Gardner},
\author[basel]{S.~Garni},
\author[glasgow,edinburgh]{D.I.~Glazier},
\author[edinburgh]{D.~Glowa},
\author[mainz]{W.~Gradl},
\author[basel]{M.~G\"unther},
\author[moscow]{G.M.~Gurevich},
\author[glasgow]{D.~Hamilton},
\author[allison]{D.~Hornidge},
\author[regina]{G.M.~Huber},
\author[basel]{A.~K{\"a}ser},
\author[mainz]{V.L.~Kashevarov},
\author[edinburgh]{S.~Kay},
\author[basel]{I.~Keshelashvili\thanksref{2}},
\author[moscow]{R.~Kondratiev},
\author[zagreb]{M.~Korolija},
\author[basel]{B.~Krusche}\ead{bernd.krusche@unibas.ch},
\author[dubna]{A.B.~Lazarev},
\author[mainz]{J.M.~Linturi},
\author[moscow]{V.~Lisin},
\author[glasgow]{K.~Livingston},
\author[basel]{S.~Lutterer},
\author[glasgow]{I.J.D.~MacGregor},
\author[glasgow]{J.~Mancell},
\author[kent]{D.M.~Manley},
\author[mainz,allison]{P.P.~Martel},
\author[giessen]{V.~Metag},
\author[bochum]{W.~Meyer},
\author[amherst]{R.~Miskimen},
\author[mainz]{E.~Mornacchi},
\author[moscow,amherst]{A.~Mushkarenkov},
\author[dubna]{A.B.~Neganov},
\author[mainz]{A.~Neiser},
\author[basel]{M.~Oberle},
\author[mainz]{M.~Ostrick},
\author[mainz]{P.B.~Otte},
\author[regina]{ D.~Paudyal},
\author[pavia]{P.~Pedroni},
\author[moscow]{A.~Polonski},
\author[mainz,ucla]{S.N.~Prakhov},
\author[amherst]{A.~Rajabi},
\author[bochum]{G.~Reicherz},
\author[jerusalem]{G.~Ron},
\author[basel]{T.~Rostomyan\thanksref{3}},
\author[halifax]{A.~Sarty},
\author[mainz]{C.~Sfienti},
\author[edinburgh]{M.H.~Sikora},
\author[mainz,gdwu]{V.~Sokhoyan},
\author[hiskp]{K.~Spieker},
\author[mainz]{O.~Steffen},
\author[gdwu]{I.I.~Strakovsky},
\author[basel]{Th.~Strub},
\author[zagreb]{I.~Supek},
\author[hiskp]{A.~Thiel},
\author[mainz]{M.~Thiel},
\author[mainz]{A.~Thomas},
\author[mainz]{M.~Unverzagt},
\author[dubna]{Yu.A.~Usov},
\author[mainz]{S.~Wagner},
\author[basel]{N.K.~Walford}
\author[edinburgh]{D.P.~Watts},
\author[basel,glasgow]{D.~Werthm\"uller},
\author[mainz]{J.~Wettig},
\author[mainz]{M.~Wolfes},
\author[edinburgh]{L.~Zana}

\address[basel]{Department of Physics, University of Basel, Basel, Switzerland}
\address[mainz]{Institut f\"ur Kernphysik, University of Mainz, Mainz, Germany}
\address[hiskp]{Helmholtz-Institut f\"ur Strahlen- und Kernphysik, University of Bonn, Bonn, Germany}
\address[regina]{University of Regina, Regina, SK S4S 0A2 Canada}
\address[kent]{Kent State University, Kent, OH, USA}
\address[glasgow]{SUPA School of Physics and Astronomy, University of Glasgow, Glasgow, G12 8QQ, UK}
\address[edinburgh]{SUPA School of Physics, University of Edinburgh, Edinburgh EEH9 3JZ, UK}
\address[dubna]{Joint Institute for Nuclear Research,141980 Dubna, Russia}
\address[pavia]{INFN Sezione di Pavia, Pavia, Italy}
\address[gdwu]{Center for Nuclear Studies, The George Washington University, Washington, DC, USA}
\address[halifax]{Department of Astronomy and Physics, Saint Marys University, Halifax, Canada}
\address[moscow]{Institute for Nuclear Research, Moscow, Russia}
\address[allison]{Mount Allison University, Sackville, New Brunswick E4L 1E6, Canada}
\address[zagreb]{Rudjer Boskovic Institute, Zagreb, Croatia}
\address[giessen]{II. Physikalisches Institut, University of Giessen, Germany}
\address[bochum]{Institut f\"ur Experimentalphysik, Ruhr Universit\"at, 44780 Bochum, Germany}
\address[amherst]{University of Massachusetts Amherst, Amherst, Massachusetts 01003, USA}
\address[ucla]{University of California at Los Angeles, Los Angeles, CA, USA}
\address[jerusalem]{Racah Institute of Physics, Hebrew University of Jerusalem, Jerusalem 91904, Israel}

\thanks[1] {also at Dipartimento di Fisica, Universit\`a di Pavia, Pavia, Italy.}
\thanks[2] {Now at Institut f\"ur Kernphysik, FZ J\"ulich, 52425 J\"ulich, Germany}
\thanks[3] {Now at Department of Physics and Astronomy., Rutgers University, Piscataway, New Jersey, 08854-8019, USA}

\begin{abstract}
The double-polarization observable $E$ and the helicity-dependent cross sections $\sigma_{1/2}$ 
and $\sigma_{3/2}$ have been measured for the first time for single $\pi^{0}$ photoproduction from 
protons and neutrons bound in the deuteron at the electron accelerator facility MAMI in Mainz, Germany. 
The experiment used a circularly polarized photon beam and a longitudinally polarized deuterated 
butanol target. The reaction products, recoil nucleons and decay photons from the $\pi^0$ meson 
were detected with the Crystal Ball and TAPS electromagnetic calorimeters. Effects from nuclear 
Fermi motion were removed by a kinematic reconstruction of the $\pi^{0}N$ final state. A comparison 
to data measured with a free proton target showed that the absolute scale of the cross sections is 
significantly modified by nuclear final-state interaction (FSI) effects. However, there is no 
significant effect on the asymmetry $E$ since the $\sigma_{1/2}$ and $\sigma_{3/2}$ components appear 
to be influenced in a similar way. Thus, the best approximation of the two helicity-dependent cross 
sections for the free neutron is obtained by combining the asymmetry $E$ measured with quasi-free 
neutrons and the unpolarized cross section corrected for FSI effects under the  assumption that the 
FSI effects are similar for neutrons and protons. 
\end{abstract}

\end{frontmatter}

\section{Introduction}

In general, the excitation spectrum of a composite system reflects the properties of the underlying
interaction. A study of the properties of nucleon resonances is as important for the 
understanding of the strong interaction as the interpretation of atomic level schemes was in the
development of Quantum Electrodynamics (QED). One difference between QED and quantum chromodynamics 
(QCD) is that, in the low-energy range of excited nucleon states, QCD cannot be solved in a 
perturbative way. Therefore,  the interpretation of experimental data commonly relied on 
phenomenological constituent quark models. However, much progress has been made with the numerical 
methods of lattice gauge calculations. While most published results have been for predictions of 
ground-state properties, the first unquenched lattice simulations of excited states have 
recently been reported \cite{Edwards_11}.

However, the comparison of the predicted nucleon excitation schemes to experimental data is still 
unsatisfactory. Only a small fraction of the predicted states has been observed so far and for most of
them, many properties such as decay branching ratios, are not well defined. 

On the experimental side, an ambitious program to investigate this spectrum with photoproduction and 
electroproduction of mesons off the nucleon has been initiated and pursued at modern electron accelerators 
like CEBAF in Newport News, MAMI in Mainz, ELSA in Bonn, ESRF in Grenoble, LEPS in Osaka, and ELPH in Sendai. 
Central to this experimental program are the measurements of several observables (in the ideal 
case a `complete' sample \cite{Chiang_97}) for different final states 
($N\pi$, $N\eta$, $N\eta '$, $N\omega$, $N\rho$, $N\pi\pi$, $N\pi\eta$, $\Sigma K$, $\Lambda K$,...).    
This program aims at coupled-channel analyses, which will result in much better constrained 
partial-wave analyses of the data. Most of the experiments measured not only differential cross sections, 
but also single and double polarization observables, which are essential for such analyses. The first 
impact of this program is visible in the {\it Review of Particle Physics} (RPP) \cite{PDG_16}, in which  
excited-nucleon states are now listed for which the experimental evidence comes only or mainly from the 
photon-induced reactions, while previously all such states were based on data from pion-induced reactions.

However, there is one aspect that is so far only addressed by few experimental data. Important 
information about the structure of $I=1/2$ $N^{\star}$ nucleon resonances is related to their isospin 
dependence of the electromagnetic excitation. The extraction of the isospin composition of the reaction
amplitudes requires data from target neutrons, which are only available as quasi-free particles bound in 
light nuclei such as the deuteron. This complicates the experiments because of the coincident detection of 
recoil neutrons and also the interpretation of the data due to Fermi smearing and FSI effects. 
In particular, the measurement of photoproduction of neutral mesons off the neutron is difficult 
for most experiments, but the measurement of the fully neutral channels is important because they are least 
effected by non-resonant backgrounds. Summaries of recent results for this program are given in 
\cite{Krusche_11,Krusche_15}.

The Crystal Ball/TAPS calorimeter at the Mainz MAMI accelerator is one of the few facilities where such
experiments can be done. Differential cross sections for neutron targets have been recently
measured with this detector
\cite{Werthmueller_13,Witthauer_13,Werthmueller_14,Werthmueller_15,Dieterle_14,Dieterle_15,Kaeser_15,Kaeser_16}, 
with the Bonn CBELSA/TAPS experiment \cite{Jaegle_08,Jaegle_11a,Jaegle_11b,Dietz_15}, and the GRAAL experiment
\cite{Kuznetsov_07} for the final states 
$n\eta$ \cite{Werthmueller_13,Witthauer_13,Werthmueller_14,Werthmueller_15,Jaegle_08,Jaegle_11a,Kuznetsov_07},
$n\pi^0$ \cite{Dieterle_14}, $n\eta '$ \cite{Jaegle_11b}, $n\omega$ \cite{Dietz_15}, $n\pi^0\pi^0$ \cite{Dieterle_15}, 
and $n\pi^0\eta$, $p\pi^-\eta$ \cite{Kaeser_15,Kaeser_16}. Single polarization observables have been measured 
for $n\eta$ \cite{Fantini_08} at GRAAL and at MAMI for $n\pi^0\pi^0$ \cite{Oberle_13}, $p\pi^-\pi^0$ \cite{Oberle_14}, 
and $n\pi^0\eta$, $p\pi^-\eta$ \cite{Kaeser_16}. Examples of unexpected results are the narrow structure in the 
excitation function of $n\eta$
\cite{Werthmueller_13,Witthauer_13,Werthmueller_14,Werthmueller_15,Jaegle_08,Jaegle_11a,Kuznetsov_07}
and the behavior of the $\gamma n\rightarrow n\pi^0$ reaction \cite{Dieterle_14}. 

The reaction amplitudes for pion
production are linear combinations of the three isospin amplitudes $A^{IS}$ (isoscalar), $A^{IV}$ (isovector),
and $A^{V3}$ (total isospin changing) \cite{Krusche_03}. Since there are four different final states 
($p\pi^0$, $n\pi^+$, $p\pi^-$, $n\pi^0$), one could argue that it is not necessary to measure the $n\pi^0$
final state. However, although plenty of cross-section data are available for the other three reactions,
the first measurement of the $\gamma n\rightarrow n\pi^0$ reaction \cite{Dieterle_14} produced results that 
did not agree with predictions based on the other isospin channels.  

Previous measurements for $\pi^0$ production have concentrated on reactions off the free proton. 
Precise angular distributions and polarization observables were reported in
\cite{Bartholomy_05,Bartalini_05,vanPee_07,Dugger_07,Elsner_09,Sparks_10,Crede_11,Thiel_12,Gottschall_14,Sikora_14}.
Until now, data for $\pi^0$ production involving neutrons were only available for inclusive measurements off
the deuteron \cite{Krusche_99}, the beam asymmetry for $\gamma n\rightarrow n\pi^0$ \cite{DiSalvo_09}
and the cross section data from \cite{Dieterle_14}.    

In the same way as for the free proton, reliable partial-wave analyses of the reaction on the neutron  
require results of further observables, namely for polarization degrees of freedom. The first data for a 
double-polarization observable for meson photoproduction off neutrons were measured at MAMI \cite{Witthauer_16a}. 
The $E$ asymmetry was extracted for several reactions using a deuterated butanol 
target and a circularly polarized photon beam. This observable allows the separation of the unpolarized 
cross section into the helicity-1/2 and helicity-3/2 parts. Results for $\eta$ production became recently 
available \cite{Witthauer_16a,Witthauer_16b,Witthauer_16c} and this paper summarizes the first results for 
this observable for single $\pi^0$ production off the neutron.

\section{Polarization observable $E$ and helicity dependent cross sections $\sigma_{1/2}$ and $\sigma_{3/2}$}
 
For a circularly polarized photon beam of polarization $P_{\odot}$ and a longitudinally polarized target of polarization 
$P_{T}$, the beam-target spin configuration can either be parallel $(\uparrow\uparrow)$ or antiparallel 
$(\uparrow\downarrow)$. The double-polarization observable $E$ and the helicity-dependent cross sections $\sigma_{1/2}$ 
and $\sigma_{3/2}$  are related by: 
\begin{equation}
E=\frac{\sigma_{1/2}-\sigma_{3/2}}{\sigma_{1/2}+\sigma_{3/2}}=
\frac{1}{P_{\odot}P_{T}}\cdot\frac{N_{1/2}-N_{3/2}}{(N_{1/2}-N_B)+(N_{3/2}-N_B)}~.
\label{eq:e}
\end{equation}
In the second part of the equation, the asymmetry is expressed in terms of the count rates $N_{1/2}$ and $N_{3/2}$ 
measured for the $(\uparrow\downarrow)$ and $(\uparrow\uparrow)$ configurations, where the polarization degrees 
$P_{\odot}$ and $P_{T}$ take into account the fraction of unpolarized photons and deuterons, respectively.
Molecular hydrogen (deuterium) cannot be polarized. For the present experiment solid butanol was used as target
material. Therefore, the count rate $N_B$ from reactions with unpolarized nucleons in the carbon and oxygen nuclei
of the butanol molecules, which is identical for both spin configurations, must be subtracted from the sum of the 
two count rates, while it cancels in the difference. This was done, as discussed below, by comparing 
the normalized yields from a butanol, a carbon foam, and a liquid deuterium target. 

The extraction of $E$, $\sigma_{1/2}$ and $\sigma_{3/2}$ can be done in different ways, which have correlated 
statistical uncertainties but different systematic ones. In the first step, the numerator and denominator 
of the left-hand part of Eq.~\ref{eq:e} are determined as:
\begin{equation}
\begin{aligned}
\sigma_{\text{diff}} & = & \sigma_{1/2}-\sigma_{3/2}\\
\sigma_{\text{sum}}  & = & \sigma_{1/2}+\sigma_{3/2},
\end{aligned}
\label{eq:ds}
\end{equation}
where for $\sigma_{\rm sum}$, the contribution from the unpolarized nuclear background must be subtracted. 
The asymmetry $E$ can then be calculated from $\sigma_{\rm diff}$ and $\sigma_{\rm sum}$ (which is labeled version (1)) 
or $\sigma_{\rm sum}$ can be replaced by $2\sigma_0$ (version (2)), where $\sigma_0$ is the unpolarized cross 
section measured with a liquid deuterium target taken from \cite{Dieterle_14}:
\begin{equation}
E^{\text{v1}} = \frac{\sigma_{\text{diff}}}{\sigma_{\text{sum}}},
                 \quad E^{\text{v2}} = \frac{\sigma_{\text{diff}}}{2\cdot\sigma_0}\, . 
\label{eq:e12}
\end{equation}
The first method has the advantage that systematic uncertainties from target densities, target geometry, photon fluxes, 
and detection efficiencies cancel in the ratio. The second method avoids the systematic effects from the subtraction 
of the unpolarized nuclear background. 
The helicity-dependent cross sections follow then from:
\begin{equation}
\begin{aligned}
\sigma^{\rm v1,v2}_{1/2} &= \sigma_0\cdot(1+E^{\rm v1,v2})\\
\sigma^{\rm v1,v2}_{3/2} &= \sigma_0\cdot(1-E^{\rm v1,v2})~,
\end{aligned}
\label{eq:s1}
\end{equation}
where $\sigma_0$ is taken in both cases from the measurement with the unpolarized deuterium target \cite{Dieterle_14}.  
         
\section{Experimental Setup}
The measurements were performed at the electron accelerator facility MAMI in Mainz, Germany. In total, four 
different beam time periods were analyzed. The asymmetry for the $\gamma n\rightarrow n\eta$ reaction
was analyzed from the same data sample and details about the experimental setup are given in \cite{Witthauer_16c}.
Therefore, we summarize here only the most important features. Longitudinally polarized electrons were
generated with optical pumping of a gallium-arsenide-phosphor (GaAsP) photocathode \cite{Aulenbacher_97} and 
accelerated in the successive accelerator stages of MAMI \cite{Kaiser_08} to $E_{e^-}\approx 1.6$~GeV. The electron 
beam impinged on a ferromagnetic Vacoflux50 foil (10 $\mu$m thickness), where it produced bremsstrahlung photons. 
The photons were energy tagged with the Glasgow spectrometer \cite{McGeorge_08} with a typical resolution of 4~MeV, 
which corresponds to around half the width of the 353 plastic scintillator bars in the focal plane of the dipole magnet. 
The detector can cover 5 - 93\% of the incident electron energies; however, since the count rates for high electron 
energies (corresponding to low photon energies) would have limited counting statistics, a part of it was deactivated 
so that only photons with energies in the range $E_{\gamma}\approx400$ - 1450~MeV were tagged. 

The electron polarization was measured with a Mott polarimeter close to the electron source and monitored  
with a M$\o$ller polarimeter viewing the bremsstrahlung foil. Both results were in good agreement and the average 
electron polarization degree was $P_{e^{-}}\approx 83\%$. The transfer of the longitudinal polarization of the electrons 
$P_{e^{-}}$ to the circular polarization of photons $P_{\gamma}$ is given by \cite{Olsen_59}:
\begin{equation}
P_{\gamma}=P_{e^{-}}\cdot\frac{4x-x^{2}}{4-4x+3x^{2}}~,\label{eq:olsen}
\end{equation}
where $x=E_{\gamma}/E_{e^-}$. 

The photon beam was defined downstream of the radiator by a collimator with 2~mm diameter, resulting in a 
beam spot size of 9~mm on the production target, which was a longitudinally polarized, frozen-spin target 
\cite{Rohlof_04}. The target had a diameter of 19.8~mm and a length of 20~mm. It was filled with deuterated butanol 
(C$_{4}$D$_{9}$OD) beads of an average diameter of 1.88~mm. Dynamic Nuclear Polarization (DNP) \cite{Bradtke_99} 
was used to polarize the deuterated butanol in a magnetic field of 1.5 T and at a temperature of 25 mK. 
For the measurement, the polarizing magnet was replaced by a small solenoidal holding coil with a magnetic 
field of 0.6 T. Relaxation times of more than 2000 h were achieved. 

\begin{figure}[!htb]
\centering
\includegraphics[width=\columnwidth]{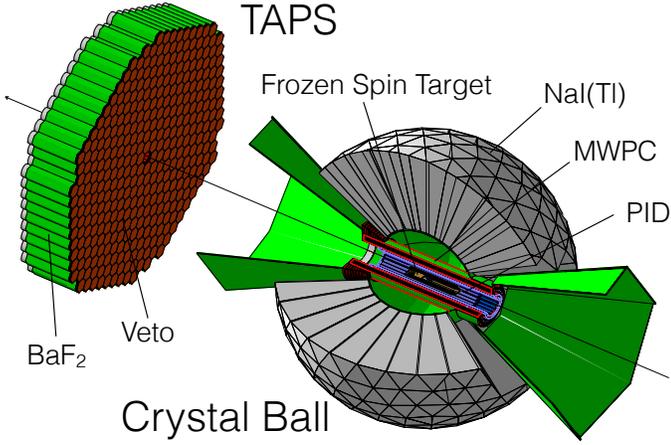}
\caption{Main detector setup of the A2 experiment at MAMI.}
\label{fig:exp}
\end{figure}

The target polarization was measured with an NMR system before and after data taking and interpolated
exponentially at times between. Typical polarization degrees were about 60\%. However, for the first 
three beam times there was a problem in the absolute determination of the polarization. This was caused by 
small field inhomogeneities ($\Delta B\leq 1.78$ mT) of the polarizing magnet. As a consequence, the polarization 
varied over the target diameter so that the measured overall polarization was not identical with the polarization 
in the target area hit by the beam. The problem was discovered \cite{Witthauer_16a,Witthauer_16c} because it is 
known that for $\eta$ production in the $N(1535)1/2^-$ resonance range, the asymmetry is close to unity. 
The butanol was chemically doped with substances with highly polarizable paramagnetic centers for efficient DNP. 
The first beam times used trityl-radicals Finland D36, which help to produce high polarization, but have a very 
narrow NMR resonance. An additional beam time used a less sensitive radical (Tempo) instead, which results in 
lower polarization degrees, but is not sensitive to small field inhomogeneities. The absolute scale of all 
asymmetries was renormalized to this beam time and it was verified that this measurement produced correct 
results for $\eta$ production \cite{Witthauer_16a,Witthauer_16c}.

In order to eliminate the background from the carbon and oxygen nuclei in the butanol, data from measurements
with a solid carbon and a liquid deuterium target were included in the analysis. The carbon target was made from a  
foam that had the same density as the heavy nuclei in the butanol target. It was fitted into the Teflon container
of the butanol so that it had the same geometry and surrounding materials. For liquid deuterium, a subsample of the 
data from \cite{Dieterle_14} was used (only those data that were measured with a target of 3 cm length and 
the same trigger conditions as for the butanol target). 

The experimental setup was identical to the one used for the measurements with the unpolarized liquid 
deuterium target and is shown in Fig. \ref{fig:exp}. Details are given in \cite{Werthmueller_14,Oberle_14}. 
An electromagnetic calorimeter, combining the Crystal Ball (CB) and the TAPS detectors, covered almost the
full solid angle around the target. The CB, constructed from 672 NaI(Tl) crystals, surrounding the target 
in a ball-like shape, covering the full azimuthal angle for polar angles between $20^{\circ}$ and $160^{\circ}$
\cite{Starostin_01}. A cylindrically shaped charged-particle identification detector (PID) \cite{Watts_05}, 
consisting of 24 plastic scintillators, was mounted inside the CB around the target. 
The setup was completed at forward angles from $5^{\circ}$ to $21^{\circ}$ by the TAPS detector 
\cite{Gabler_94} comprised of 384 hexagonally shaped BaF$_{2}$ crystals arranged in 11 rings.
Each of the 18 crystals of the two inner-most rings were replaced by four PbWO$_{4}$ crystals
to achieve higher rate capability at the forwardmost angles. A 5 mm thick plastic scintillator was mounted 
in front of each BaF$_{2}$ crystal for charged particle identification (CPV detector).

For the generation of the trigger, TAPS was divided into six triangular logic sectors and the CB
into sectors of 16 adjacent crystals. A TAPS sector contributed to the trigger when at least
one module had an energy deposition above 35~MeV and the CB sectors contributed for energy depositions
between 10 - 30 MeV. A total sector multiplicity of two was required for the trigger and in
addition, a minimum energy deposition (analog sum of the energy signals) of 250~MeV in the CB 
detector. Events from $\pi^0$ decays with both photons in TAPS were not included in the
trigger.

\section{Data Analysis}

The analysis was analogous to that of the data for $\eta$ photoproduction \cite{Witthauer_16c} from
the same beam-time periods and, apart from the treatment of the nuclear background, followed the strategies 
developed for the unpolarized data from several reactions discussed in detail in 
\cite{Werthmueller_13,Witthauer_13,Werthmueller_14,Dieterle_14,Dieterle_15,Kaeser_15,Kaeser_16,Oberle_13,Oberle_14}.
Therefore, only a short summary is given here. 

The data from the butanol target were analyzed together with data from the carbon foam and liquid deuterium targets
to allow for the elimination of the unpolarized background for analysis version (1). The analysis strategies and 
cuts were tuned for the liquid deuterium data and then applied identically to the butanol and carbon data. 

In the initial step of the analysis, all modules were calibrated for their energy and timing
response. Subsequently, hits in the calorimeters were assigned to particle types such as photons, neutrons, and 
protons. For hits in the CB, an $E-\Delta E$ analysis using the PID was performed. For hits in TAPS, pulse-shape 
analysis (PSA), time-of-flight versus energy analysis, and the response of the CPV detector were used. 
These procedures are described in detail in 
\cite{Werthmueller_13,Witthauer_13,Werthmueller_14,Dieterle_14,Dieterle_15,Kaeser_15,Kaeser_16,Oberle_13,Oberle_14}.
The only remaining ambiguity was that photons and neutrons in the CB cannot be distinguished event-by-event
(see e.g. \cite{Witthauer_13,Werthmueller_14}).

Events with exactly three neutral (candidates for $n\pi^0$ final state) or two neutral and one charged 
(candidates for $p\pi^0$ final state) hits were accepted for further analysis. For the latter class of events, the 
assignment of hits to $\pi^0$ decay photons and to the recoil proton was straightforward. For the events with three
neutral hits, where ambiguities between photons and neutrons occurred, a $\chi^2$ test was performed to identify the 
recoil neutron. This procedure is described in detail in previous publications
\cite{Werthmueller_13,Witthauer_13,Werthmueller_14,Dieterle_14,Dieterle_15,Kaeser_15,Kaeser_16,Oberle_13,Oberle_14}.
This test compared the invariant mass of the two photon candidates $m_{\gamma_{1}\gamma_{2}}$ to the nominal 
pion mass $m_{\pi^{0}}$:
\begin{equation}
\chi^{2} = \left(\frac{m_{\gamma_{1}\gamma_{2}}-m_{\pi^{0}}}
{\Delta m_{\gamma_{1}\gamma_{2}}}\right)^{2}~,
\label{eq:chi2}
\end{equation}
where the resolution $\Delta m_{\gamma_{1}\gamma_{2}}$ was determined from the simulated detector response. 
For each event with three neutral hits, the combination with the lowest $\chi^2$ was selected for the $\pi^0$ 
decay photons and the remaining neutral hit as a neutron candidate.

The next step in the analysis was the clean identification of the $\gamma N\rightarrow N\pi^0$ reaction.
Background originates mainly from $\pi^0\pi$, e.g. when for $\pi^0\pi^{\pm}$ pairs the charged pion was not 
detected or when for $\pi^0\pi^0$ pairs one photon was lost and the other was falsely assigned as a neutron. 
This reaction identification was based on the analysis of the reaction kinematics. It also reduced the 
unpolarized background from the carbon nuclei in the butanol, which due to Fermi smearing and FSI effects, 
is less likely to pass these selection cuts. 

The first condition was based on the coplanarity of the reconstructed $\pi^{0}$ and nucleon. Due to 
momentum conservation, the difference in the azimuthal angle $\Delta\phi$ between the $\pi^0$ and the recoil
nucleon must be 180$^{\circ}$ and this is normally not the case when additional particles have escaped detection.
Examples for such analyses for different reactions are shown in  
\cite{Witthauer_13,Werthmueller_14,Dieterle_15,Kaeser_15,Oberle_14,Witthauer_16c}; however, since 
undetected low-energy pions carry only small momenta, this condition is not stringent.  

A more effective constraint on the reaction kinematics is the missing mass. For this analysis, the recoil nucleon 
was treated as a missing particle (although it was detected) and its mass was calculated from
the four momenta of the incident photon $P_{\gamma}$, the initial-state nucleon $P_{N}$, and the final-state pion 
$P_{\pi^0}$:  
\begin{equation}
\Delta M = \left|P_{\gamma} + P_{N} - P_{\pi^0}\right| - m_{N}~,
\label{eq:mm}
\end{equation} 
where the nucleon mass $m_{N}$ was subtracted so that true $\gamma N\rightarrow N\pi^0$ events were expected at
$\Delta M = 0$. The signal was broadened by Fermi motion because the initial-state nucleon was not at rest.

\begin{figure}[!thb]
\centerline{
\resizebox{0.48\textwidth}{!}{\includegraphics{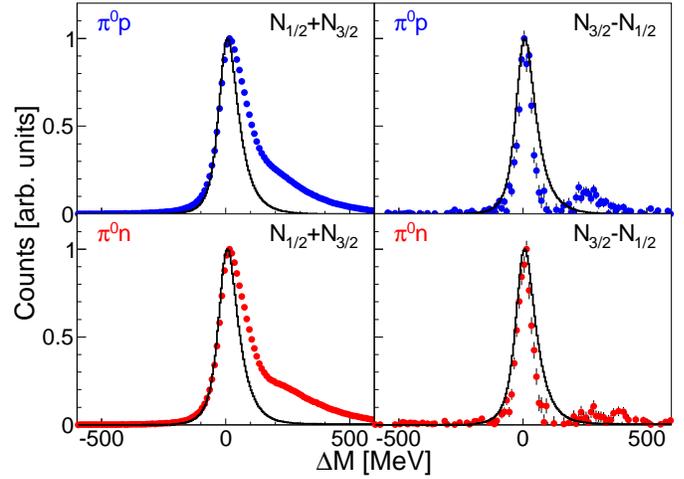}}}
\caption{Missing-mass spectra for the sum and difference of the two polarization states for the $p\pi^0$ and
$n\pi^0$ final states measured with the butanol target. Coplanarity and invariant mass cuts were applied. 
The solid lines represent MC simulations for an unpolarized liquid deuterium target.}
\label{fig:missmas_diff}
\end{figure}

Figure~\ref{fig:missmas_diff} shows the missing-mass spectra for the sum and the difference of the two helicity
components. The nuclear background in the sum is clearly visible. However, in the difference only a small, well
separated background from double pion production contributes (the nuclear contributions cancel and the asymmetry for
background production off the deuteron seems to be small). For analysis version (2) only this difference is
important. Also shown are Monte Carlo (MC) simulations of the line shape of the $\pi^0$ signal for a liquid deuterium 
target with the GEANT4 code \cite{Geant4}. They agree well with the difference data (small residual deviations are 
due to the energy dependence of Fermi motion and experimental resolution; the spectra are integrated over angles 
and incident photon energy).

\begin{figure*}[!thb]
\centerline{
\resizebox{0.88\textwidth}{!}{\includegraphics{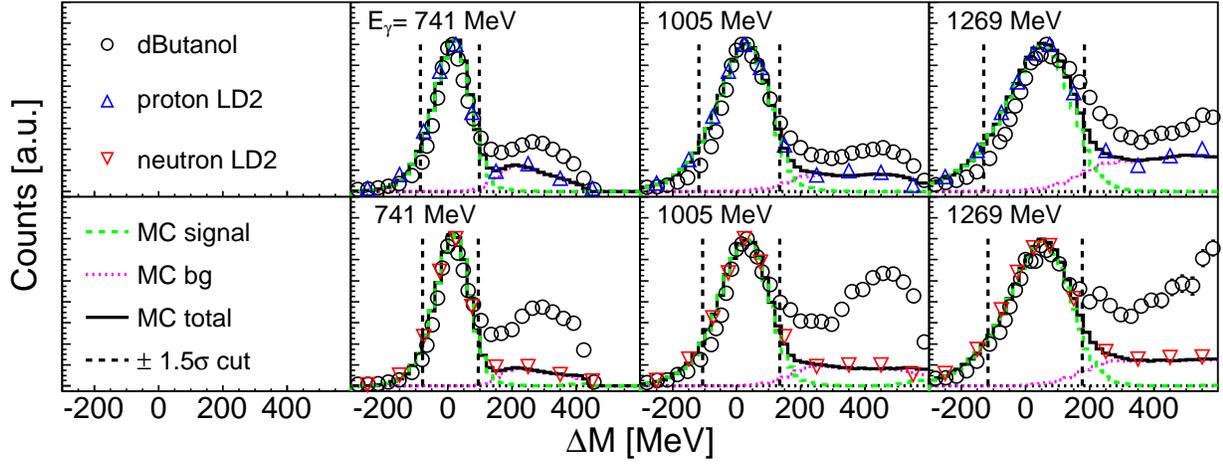}}}
\caption{Missing-mass spectra for the reaction on the quasi-free proton (upper row) and neutron 
(lower row) for three representative bins of incident photon energy. The spectra were integrated over 
the angular range $\cos(\theta_{\pi^{0}}^{\ast})=[-1.0,0.8]$ and were filled after the $\chi^{2}$ 
selection, PSA, coplanarity, and invariant-mass cuts had been applied. Monte Carlo simulations are for the
LD2 target. The notation is given in the figure.}
\label{fig:missmas}
\end{figure*}

Analysis version (1) uses also the sum spectra from the butanol target. Figure \ref{fig:missmas} shows the missing-mass 
distributions for three representative energy bins. Also shown are the results of MC simulations with GEANT4 
\cite{Geant4}. The simulations included the effects from nuclear Fermi motion in deuterium nuclei using the 
parameterization of the nucleon momentum distribution from \cite{Lacombe_81}. In addition to the signal from single 
$\pi^0$ production background from the $\pi^0\pi^0$, $\pi^0\pi^{\pm}$, $\pi^0\pi^+\pi^-$, and $\eta$ final states 
was simulated, all in coincidence with recoil protons and neutrons.
The distributions from the liquid deuterium data were fitted with the line shapes of the simulated signal and 
background contributions. The agreement between data and simulation was good for all bins. 
The applied cuts at $\pm1.5\sigma$ efficiently eliminated the background contributions, which originated mainly 
from the $\gamma N\to\pi^{0}\pi^{\pm}N$ reaction. Also shown in Fig.~\ref{fig:missmas} are the missing-mass 
distributions for the deuterated butanol target. Their shape is different (mainly in the background region) due to 
FSI effects and larger Fermi momenta. A significant fraction of background from the carbon nuclei 
was eliminated with the cuts optimized for the deuterium target; however, there is background intruding into the 
peak region that also has to be subtracted (see below).

\begin{figure*}[!htb]
\centerline{
\resizebox{0.88\textwidth}{!}{\includegraphics{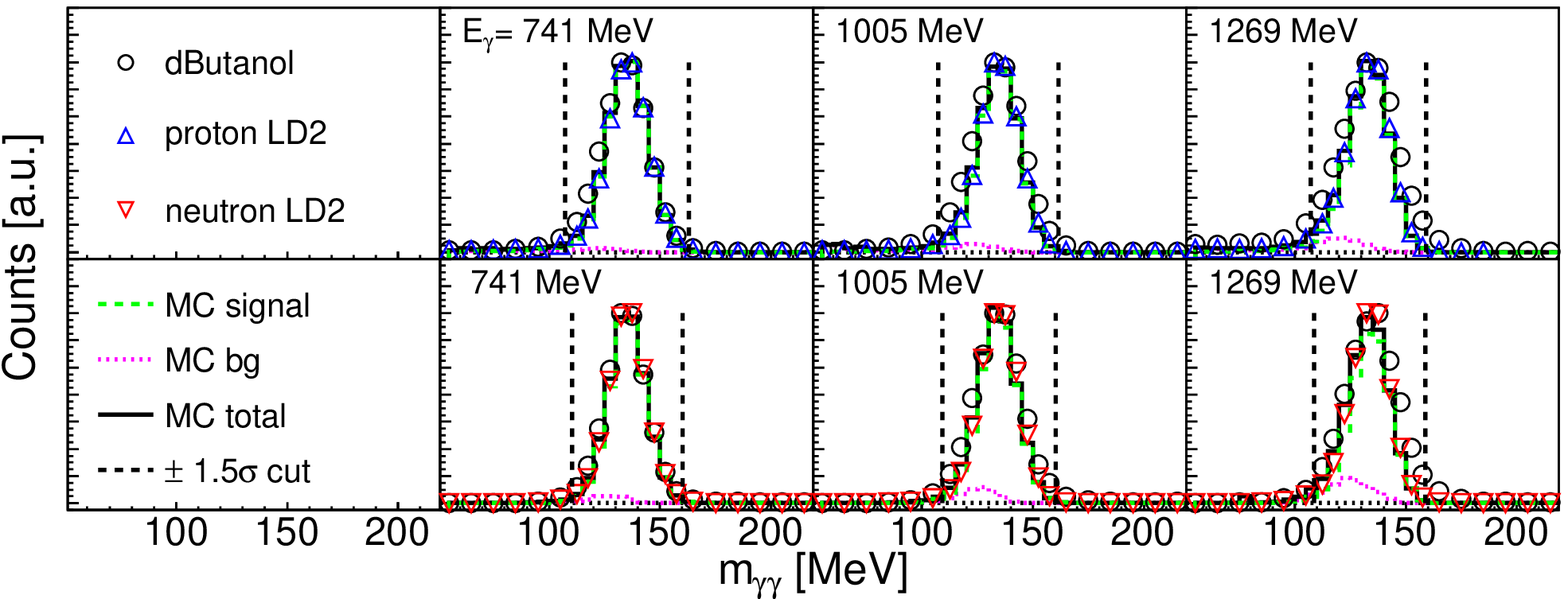}}}
\caption{Invariant-mass spectra for the reaction on the quasi-free proton (upper row) and neutron 
(lower row) for three representative bins of incident photon energy. The spectra were integrated over 
the angular range $\cos(\theta_{\pi^{0}}^{\ast})=[-1.0,0.8]$ and were filled after the $\chi^{2}$ 
selection, PSA, coplanarity, and missing-mass cuts had been applied. Monte Carlo simulations are for LD2
target. The notation is given in the figure.}
\label{fig:invmass}
\end{figure*}

In addition to the missing-mass analysis described above, an invariant-mass cut was also applied for identification
of the $\pi^0$ mesons. The invariant mass of the photon pairs was calculated from:   
\begin{equation}
m_{\gamma\gamma} = \left|P_{\gamma_{1}}+P_{\gamma_{2}}\right|,
\label{eq:im}
\end{equation}
where $P_{\gamma_{1,2}}$ are the four momenta of the two decay photons. Typical spectra are shown in 
Fig.~\ref{fig:invmass}. The spectra from liquid deuterium and butanol targets are similar because the 
invariant mass of particles is not influenced by Fermi motion or FSI effects.

\begin{figure*}[!htb]
\centerline{
\resizebox{0.88\textwidth}{!}{\includegraphics{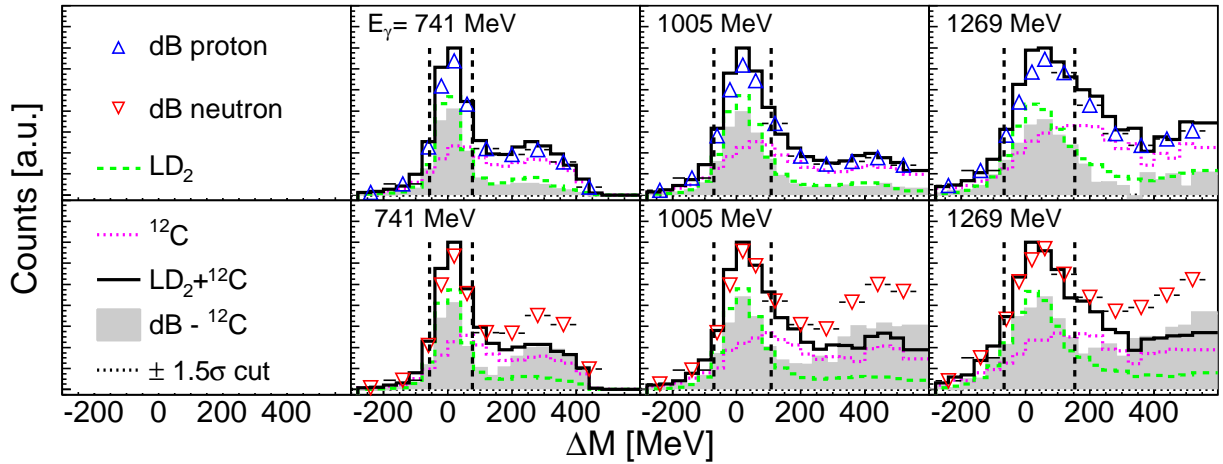}}}
\caption{Missing-mass spectra from the analysis of liquid deuterium (dashed green lines), carbon 
(dotted magenta lines), and deuterated butanol data 
(top row: coincident protons, bottom row: coincident neutrons) for three representative bins
of incident photon energy. The spectra were absolutely normalized, integrated over the angular range 
$\cos(\theta_{\pi^{0}}^{\ast})=[-1.0,0.8]$ and were filled after the $\chi^{2}$ selection, PSA, 
coplanarity, and invariant-mass cuts had been applied. The notation is given in the figure.}
\label{fig:missmas2}
\end{figure*}

The deuterated butanol data contain not only events from reactions on the polarized nucleons bound in deuterium, 
but also from unpolarized nucleons bound in the carbon and oxygen nuclei. This background can be best identified
in the missing-mass spectra. Spectra absolutely normalized by flux, target density, and detection efficiency
are compared in Fig.~\ref{fig:missmas2} for the measurements with the butanol, liquid deuterium, and solid carbon
targets. They are integrated over the angular range of cos$(\theta_{\pi^0}^{\star})$ from -1 to 0.8 because the
extreme forward angles had only marginal statistical quality due to the experimental trigger conditions.
Only the absolute scale of these spectra is arbitrary and there are no free parameters in the relative scales
of the data from different targets. Good agreement was observed between the data measured with the butanol target
and the sum of the data from liquid deuterium and carbon data in the signal region. Discrepancies occurred only 
in the background region of the spectra for the data in coincidence with neutrons. The gray shaded histograms 
indicate the difference of butanol and carbon data. These spectra were used to extract the 
carbon-subtracted yields from the butanol data for which the indicated cuts were applied.

The measured yields were absolutely normalized as for the unpolarized cross section in \cite{Dieterle_14} with the
measured photon flux, the target density, the detection efficiency, and the decay-branching ratio for the
$\pi^0\rightarrow 2\gamma$ decay \cite{PDG_16}. The determination of the photon flux was based on the counting of the
scattered electrons in the focal plane of the tagging spectrometer. The fraction of bremsstrahlung photons that 
pass the collimator and impinge on the target was measured by direct counting of the photons by a leadglass 
detector moved into the beam at a low beam intensity. The detection efficiency was determined 
mainly with the MC simulations using the GEANT4 code \cite{Geant4}. These simulations are very reliable 
for the detection of photons. Additionally, independent analyses from experimental data have been used to correct 
imperfections in the detection efficiency of the recoil nucleons in a similar way as described in \cite{Werthmueller_14}.
One should note that for the determination of $E$, those normalizations cancel to a large extent.
In version (1), they enter only via the relative normalization of the carbon background and in version (2) via the 
comparison of cross sections measured with the liquid deuterium and butanol target. However, for both 
measurements, the same kind of analysis was applied so that most systematic uncertainties cancel. 

The observables were extracted as a function of cos$(\theta_{\pi^{0}}^{\ast})$, where $\theta_{\pi^{0}}^{\ast}$
is the polar angle of the pion in the $\pi^{0}N$ center-of-mass (cm) system, and the final-state invariant mass $W$:
\begin{equation}
W=\sqrt{s} = \left|P_{N}+P_{\pi}\right|.
\label{eq:W}
\end{equation}
$P_N$ and $P_{\pi}$ are the four momenta of the pion and the recoil nucleon, respectively. The four momentum 
of the pion was defined by the decay photons measured in the calorimeter and the four momentum of the recoil 
nucleon by its measured azimuthal and polar angles and overall momentum and energy conservation 
(see \cite{Krusche_11,Werthmueller_14,Jaegle_11a,Kaeser_16,Witthauer_16c}). This analysis removed the effects 
from nuclear Fermi smearing in the reconstruction of $W$. The resolution for $W$ is then due to the 
measurement of the momenta of the decay photons and the angles of the recoil nucleon and was between
22~MeV at $W$=1.3 GeV and 60~MeV at $W$=1.8 GeV (FWHM). No systematic discrepancies between analysis versions (1)
and (2) were observed.

In this Letter, we present those results that can be compared most directly to model predictions (further results
will be summarized in an upcoming archival paper).
The results for the unpolarized differential and total cross sections for quasi-free
$\pi^0$ production discussed in \cite{Dieterle_14} show significant effects from FSI. This was investigated with
a comparison of the results for the $p\pi^0$ final state for free and quasi-free protons. The main finding was
that there is a pronounced effect on the absolute scale of the cross section, but only small effects for the
shape of the angular distributions. Larger effects were predicted by FSI models \cite{Tarasov_16} at extreme 
forward angles, which, however, were not measured either in \cite{Dieterle_14} or in the present 
experiment. However, agreement between the model results and the measured quasi-free proton data is not
good enough for quantitative corrections of the FSI effects. 

Therefore, for the results discussed in the next section, the FSI effects have been `corrected' under two
assumptions. The first assumption is that the FSI effects do not significantly depend on the helicity state. 
This means that they cancel in Eqs.~\ref{eq:e} and \ref{eq:e12} and thus $E$ is not significantly affected. 
This assumption can be tested by a comparison of the $E$ results measured here for quasi-free protons and 
the free-proton results from \cite{Gottschall_14}. The second assumption is that the FSI effects are similar 
(apart from the unmeasured extreme forward angles) for quasi-free protons and quasi-free neutrons so that the 
ratio of free to quasi-free proton data can be used to correct the quasi-free neutron results as in \cite{Dieterle_14}. 
Thus the results for the helicity-dependent cross sections $\sigma_{1/2}$ and $\sigma_{3/2}$ have been obtained 
by inserting in Eq.~\ref{eq:s1} for $\sigma_0$ for the proton, the free-proton cross section and for 
the neutron, the FSI corrected neutron cross section from \cite{Dieterle_14}. Therefore, these results can 
be compared directly to model predictions for $\pi^0$ production off free nucleons.  

\section{Results}
\begin{figure*}[!thb]
\centerline{
\resizebox{0.8\textwidth}{!}{\includegraphics{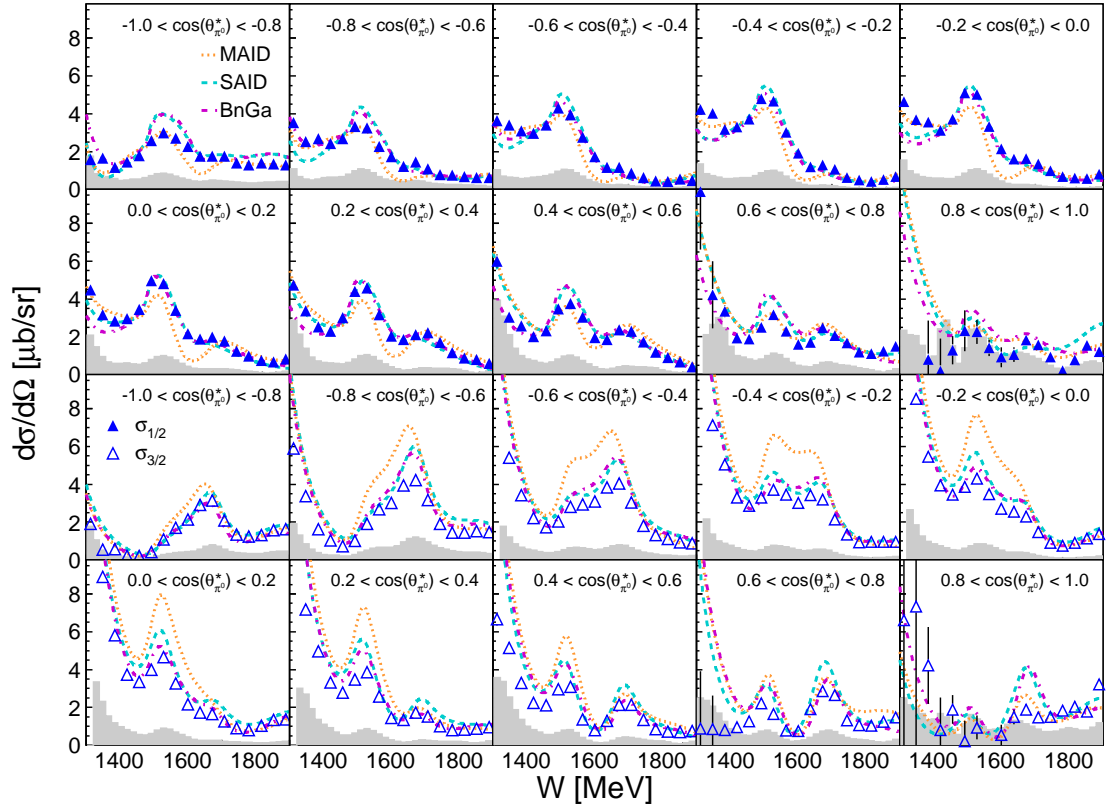}}}
\caption{Differential helicity-dependent cross sections for the quasi-free proton for different bins
of cos$(\theta_{\pi^{0}}^{\ast})$. 
Upper two rows: $\sigma_{1/2}$, lower two rows: $\sigma_{3/2}$. The gray histograms represent the 
systematic uncertainties of the results. Dotted orange lines: MAID analysis \cite{Drechsel_99}. 
Dashed green lines: SAID partial-wave analysis \cite{Workman_12}. Dashed-dotted magenta lines: 
BnGa analysis \cite{Anisovich_10}.}
\label{fig:pro_ang}
\end{figure*}

\begin{figure*}[!htb]
\centerline{
\resizebox{0.8\textwidth}{!}{\includegraphics{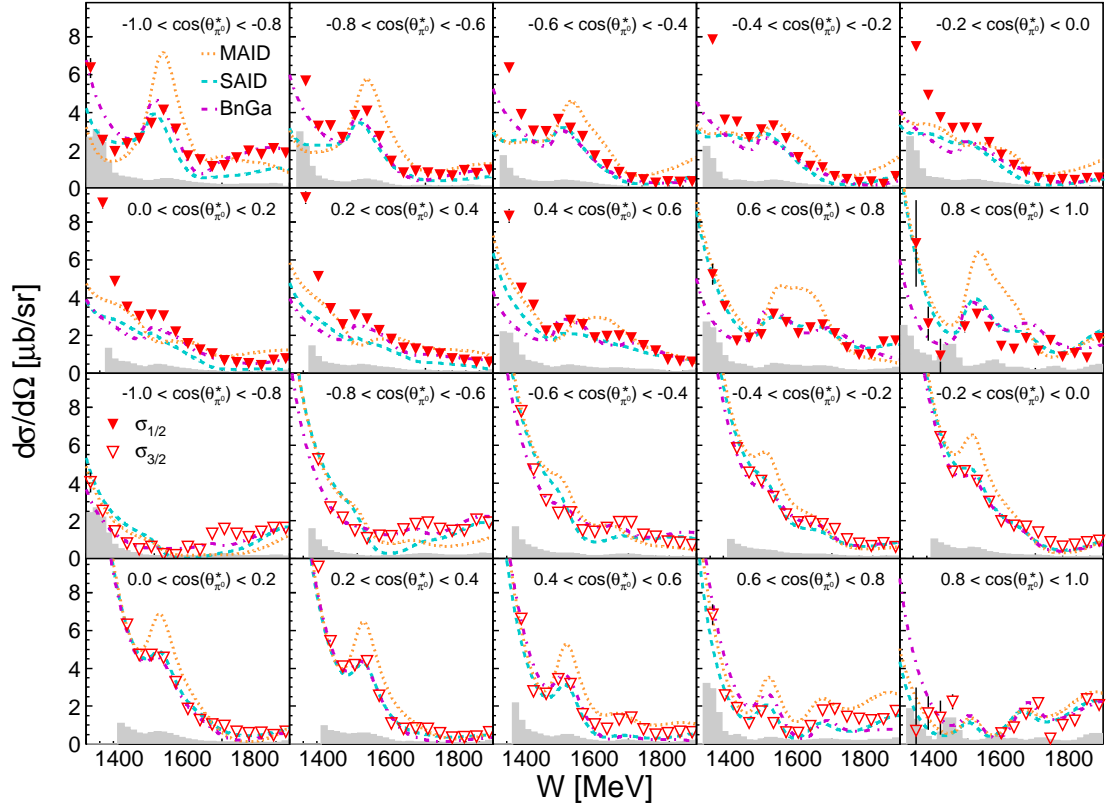}}}
\caption{Same as Fig. \ref{fig:pro_ang}, but for the quasi-free neutron.}
\label{fig:neu_ang}
\end{figure*}

The angular dependence of the helicity cross sections is summarized in Figs.~\ref{fig:pro_ang},\ref{fig:neu_ang}. 
Excitation functions are shown for different bins of the pion polar angle in the pion-nucleon
center-of-momentum (cm) frame. The results from the two extraction methods version (1), version (2) were
averaged. Statistical uncertainties were linearly averaged, because they are highly correlated (dominated by
the numerator in Eq.~\ref{eq:e} which is identical for both extraction methods). Overall, the agreement between
experimental proton data and the SAID and BnGa results is reasonably good. It is much worse for 
MAID, which is the only model that has not yet included free-proton $E$ data in fits. Also for the 
neutron, the rough features of the split into the helicity-1/2 and helicity-3/2 components of the cross 
section are in agreement with the SAID and BnGa predictions, although certainly a refit including the new 
data will be necessary to describe details.

\begin{figure}[thb]
\centerline{
\resizebox{0.48\columnwidth}{!}{\includegraphics{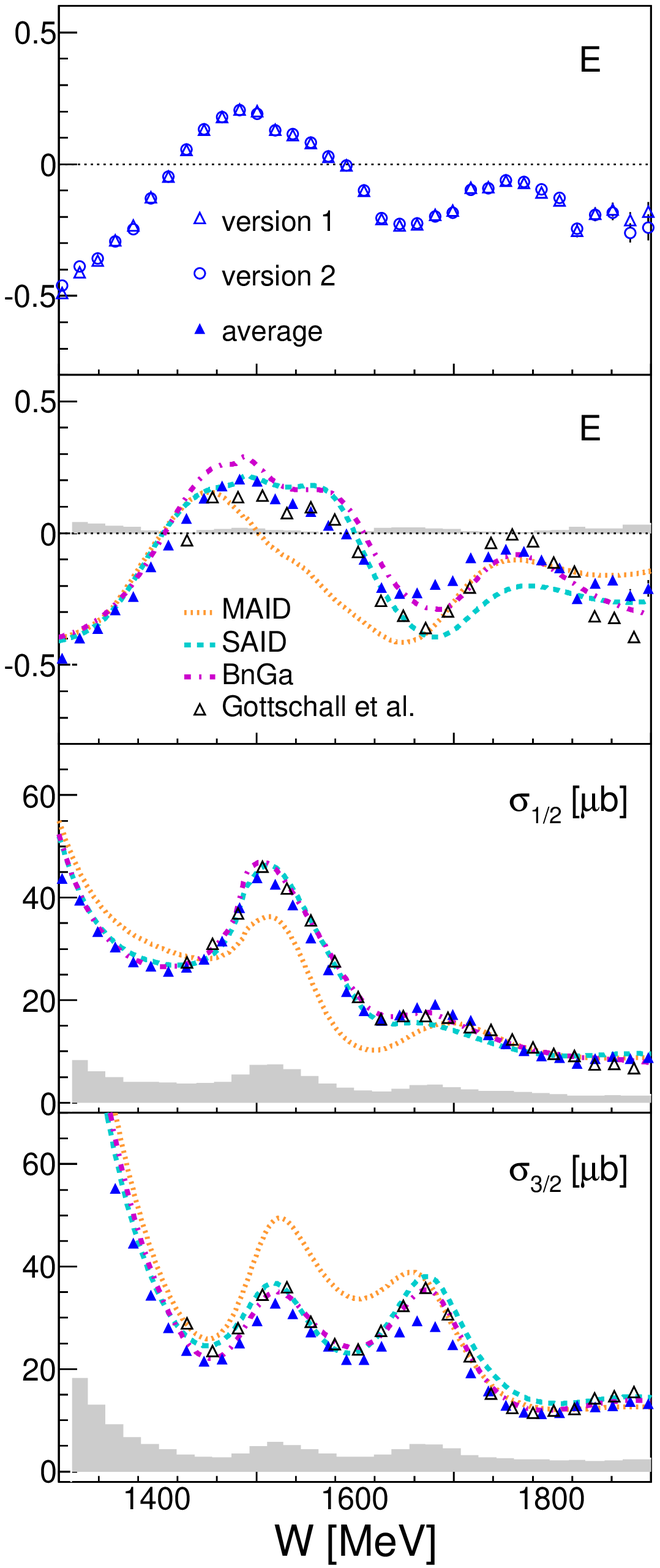}}
\resizebox{0.48\columnwidth}{!}{\includegraphics{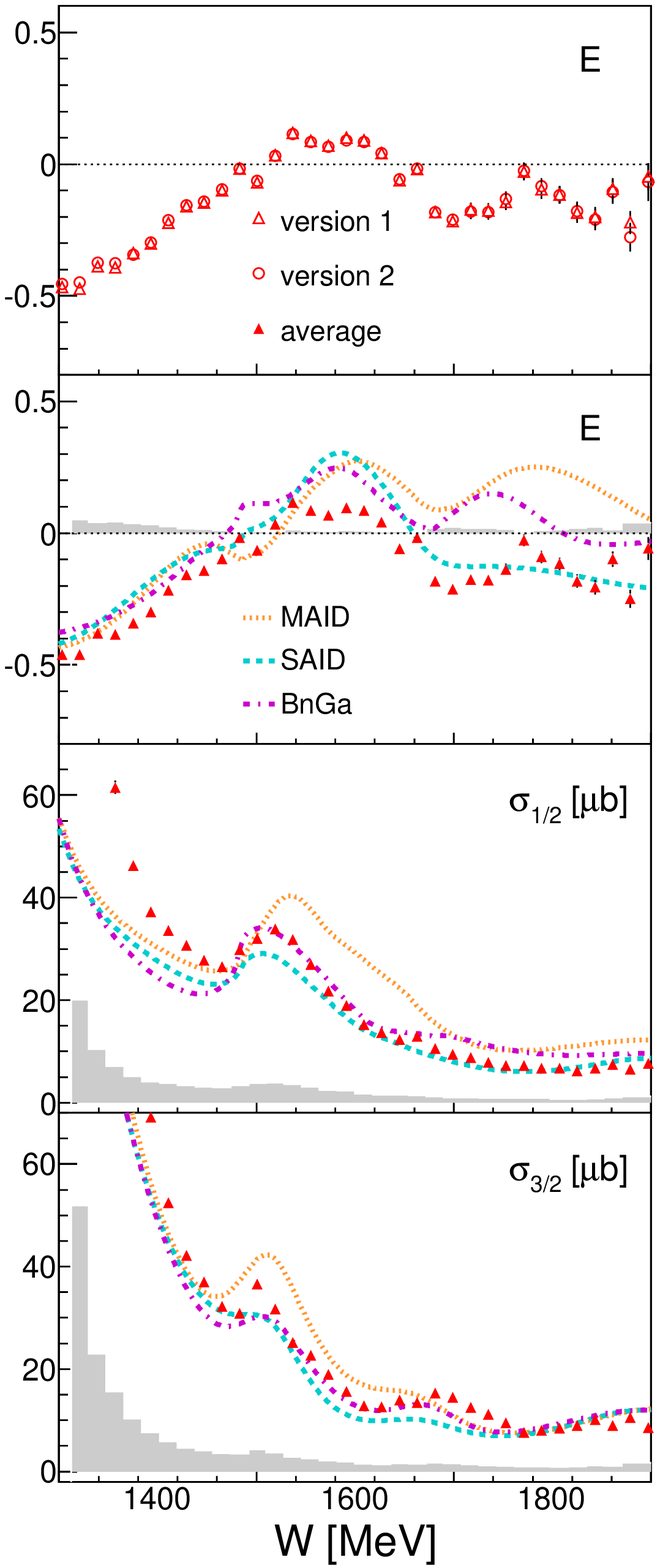}}}
\caption{Left-hand side: results for the quasi-free proton target, from top to bottom: 
(1) comparison of the results from the two analyses (1) and (2) for the asymmetry $E$, 
(2) comparison of the average of the two analyses of $E$ to results for the free proton \cite{Gottschall_14} and 
to model results from MAID \cite{Drechsel_99,Drechsel_07}, SAID \cite{Arndt_02,Workman_12}, and BnGa \cite{Anisovich_10},
(3) and (4) results for $\sigma_{1/2}$, $\sigma_{3/2}$ (see text) compared to the same model results. 
Gray histograms: systematic uncertainties. Right-hand side: same notation, but for the quasi-free neutron target.}
\label{fig:pres}
\end{figure}

The results for the angle-integrated double-polarization observable $E$ (Eqs.~\ref{eq:e},\ref{eq:e12} with total cross 
sections) and the helicity-dependent total cross sections $\sigma_{1/2}$ and $\sigma_{3/2}$ (Eq.~\ref{eq:s1})  
are summarized in Fig.~\ref{fig:pres}. 


The plots at the top compare the $E$ asymmetry for the proton and neutron obtained by either normalizing 
to the carbon-subtracted sum of the helicity states measured with the butanol target (version (1)) or to the 
unpolarized cross section measured with the liquid deuterium target \cite{Dieterle_14} (version (2)). 
The two analyses are in almost perfect agreement, indicating that there are no normalization issues or problems 
with the carbon background subtraction. Systematic uncertainties have been estimated from the uncertainty of the 
measurement of the polarization degrees (2.7\% for the photon beam, 10\% for the target) and the agreement between 
the two analyses.

The second panels from the top compare the average of the two analyses to the predictions from the MAID, SAID, 
and BnGa analyses \cite{Drechsel_07,Workman_12,Anisovich_10}. For the proton, results from a measurement with a free proton 
target \cite{Gottschall_14} are also shown. They agree well with the present results in the second resonance region.
In the third resonance region, some deviations occur due to the poorer resolution of the $W$ reconstruction at large $W$,
which broadens the peak structure. Large FSI effects had been observed for the absolute scale of 
the unpolarized cross section \cite{Dieterle_14}, but the good agreement of the asymmetry measured for free and 
quasi-free protons indicates that FSI must be very similar for both helicity components. 

The two bottom panels compare the helicity-dependent cross sections $\sigma_{1/2}$ and $\sigma_{3/2}$, averaged over 
the two versions, to model predictions and, for the proton, also to the free-proton results from \cite{Gottschall_14}.
For the proton target, SAID and BnGa results agree well with the data because free-proton data from Gottschall et al. 
\cite{Gottschall_14} had been included in their previous fits. The agreement is poorer for the neutron 
target, for which no previous experimental results exist. 

Some simple inferences can be drawn directly from the comparison of data and model predictions.
At the lowest measured $W$ values, the tail of the $\Delta(1232)3/2^{+}$ is just visible. For the proton,
this is reproduced very well by the model results, in particular those from SAID and BnGa. Here, one should 
remember that the agreement between the unpolarized cross section and the model results is trivial, 
because the models have been fitted to the free-proton data base. However, the split into $\sigma_{1/2}$ and
$\sigma_{3/2}$ is less trivial because the results for $E$ from ELSA \cite{Gottschall_14} do not cover
this energy range. Also, the neutron $\sigma_{3/2}$ data are well reproduced in this range by the SAID and BnGa
predictions. 

The low-energy neutron $\sigma_{1/2}$ data are underestimated by the models. Contributions from the $N(1440)1/2^{+}$ 
Roper resonance could be more important in this energy range for the neutron than for the proton. However, 
the systematic uncertainty of the data in this region is also substantial. 

The second resonance bump is quite well reproduced for the proton in $\sigma_{3/2}$ and also in $\sigma_{1/2}$
by SAID and BnGa. MAID, which was not fitted to the ELSA data for $E$ \cite{Gottschall_14}, has a different 
split into the two helicity components. Major contributions are from the $N(1520)3/2^{-}$ and the 
$\Delta(1700)3/2^-$ states. Also the $N(1535)1/2^{-}$ and the $N(1440)1/2^{+}$ contribute to the helicity-1/2 part. 
This structure looks different for the neutron target, in particular for the helicity-3/2 part for which 
the peak is narrower than predicted. 

The third resonance bump is only a pronounced structure for the helicity-3/2 component of the proton target.   
This is due to the contribution of the $N(1680)5/2^{+}$ state, which couples to the proton much more strongly
in $\sigma_{3/2}$ than in $\sigma_{1/2}$ ($A_{3/2}^{p}=133\pm12$, $A_{1/2}^{p}=-15\pm6$). It couples only
weakly to the neutron ($A_{3/2}^{n}=-33\pm9$, $A_{1/2}^{n}=29\pm10$) \cite{PDG_16} in both helicity states. 

\section{Summary and Conclusions}
Results for the double-polarization observable $E$ and the helicity-dependent cross sections $\sigma_{1/2}$ 
and $\sigma_{3/2}$ were measured with high statistics for photoproduction of $\pi^{0}$ mesons from 
quasi-free protons and neutrons. The results obtained with the butanol
target for the unpolarized cross section (not shown in this paper) agree with the earlier measurements with
liquid deuterium targets and confirm the large effects from nuclear final-state interactions on the absolute
scale of the cross section. However, FSI effects are much less relevant for the present results. Only the 
asymmetry $E$ was directly extracted from the quasi-free data. In the case of the proton, the free-proton
cross section and the present results for $E$ were used to construct the results for $\sigma_{1/2}$ 
and $\sigma_{3/2}$. This means that for the proton, only FSI effects on $E$ matter. Since the present results 
for $E$ and for $\sigma_{1/2}$ and $\sigma_{3/2}$ are in good agreement with a direct measurement with a liquid 
hydrogen target \cite{Gottschall_14}, FSI effects on $E$ must be small for the proton. The neutron data were
constructed in a similar way. In this case, the quasi-free results from \cite{Dieterle_14} corrected for FSI
under the assumption that it is identical for quasi-free protons and neutrons, were used. This means that also 
for the neutron, only the higher-order effects, i.e. differences in FSI between quasi-free protons and neutrons
and/or between $\sigma_{1/2}$ and $\sigma_{3/2}$ components must be considered, but not the large (up to 35\%)
effect on the absolute scaling of the cross section. It is thus reasonable to compare the data to the results
from model predictions for free nucleons. For the proton, agreement between the present data and the
results from the SAID \cite{Workman_12} and BnGa \cite{Anisovich_10} analyses are good, as expected, because
both analyses had already been fitted using similar data for the free proton. The agreement with 
MAID \cite{Drechsel_99} is clearly poorer. Description of the neutron data will obviously require refits for 
all three models.   
  
\vspace*{0.5cm}
{\bf Acknowledgments}
We wish to acknowledge the outstanding support of the accelerator group and operators of MAMI.
This work was supported by Schweizerischer Nationalfonds (200020-156983, 132799, 121781, 117601),
Deutsche For\-schungs\-ge\-mein\-schaft (SFB 443, SFB 1044, SFB/TR16), the INFN-Italy,
the European Community-Research Infrastructure Activity under FP7 programme (Hadron Physics,
grant agreement No. 227431),
the UK Science and Technology Facilities Council (ST/J000175/1, ST/G008604/1, ST/G008582/1,ST/J00006X/1, and
ST/L00478X/1),
the Natural Sciences and Engineering Research Council (NSERC, FRN: SAPPJ-2015-00023), Canada. This material
is based upon work also supported by the U.S. Department of Energy, Office of Science, Office of Nuclear
Physics Research Division, under Award Numbers DE-FG02-99-ER41110, DE-FG02-88ER40415, and DE-FG02-01-ER41194
and by the National Science Foundation, under Grant Nos. PHY-1039130 and IIA-1358175.

\end{document}